\documentclass[AMA,twocolumn]{USG} 
\usepackage{anyfontsize}

\usepackage{amsmath,amsfonts,amsthm,wasysym}
\usepackage{colortbl,rotating}
\usepackage[dvipsnames]{xcolor}
\usepackage{soul,url,hyperref}
    \hypersetup{
      colorlinks,
      linkcolor={green!80!black},
      citecolor={red!70!black},
      urlcolor={blue!70!black}
    }
\usepackage{tikz}
    \usetikzlibrary{arrows.meta,positioning}
    \def\BibTeX{{\rm B\kern-.05em{\sc i\kern-.025em b}\kern-.08em
        T\kern-.1667em\lower.7ex\hbox{E}\kern-.125emX}}
\allowdisplaybreaks

\usepackage{steinmetz}
\graphicspath{{./images/}}

\newcommand{\red}{\textcolor{Red}}

\newcommand{\green}{\textcolor{Green}}
\newcommand{\redone}{\red{0}}
\newcommand{\redzer}{\red{1}}
\newcommand{\greone}{\green{1}}
\newcommand{\grezer}{\green{0}}
\newcommand{\grebul}{\green{$\CIRCLE$}}
\newcommand{\redbul}{\red{$\CIRCLE$}}
\newcommand{\redbar}{\red{$-$}}

\newcommand{\drt}[2]{\begin{tabular}{@{}c@{}}#1\\#2\end{tabular}}
\newcommand{\drts}[2]{\begin{sideways}\begin{tabular}{@{}c@{}}#1\\#2\end{tabular}\end{sideways}}
\newcommand{\lgc}{\cellcolor{black!10}}
\newcommand{\dgc}{\cellcolor{black!30}}

\articletype{REGULAR ARTICLE}%

\received{00}
\revised{00}
\accepted{00}
\journal{System engineering}
\volume{0}
\copyyear{2026}
\startpage{1}
\articledoi{10.1002/0000}

\begin{document}

\title{Novel methodology for obtaining design structure matrices using network identification}
\transtitle{Novel methodology for obtaining design structure matrices using network identification}

\author[1,2]{E.M.M. (Lizan) Kivits}[https://orcid.org/0000-0003-1035-1878]
\author[1,2]{Matthijs {van Berkel}}[https://orcid.org/0000-0001-6574-3823]
\author[1,2]{Paulo A. Figueiredo}[https://orcid.org/0009-0004-7052-6793]
\author[1,3]{Marco R. {de Baar}}[https://orcid.org/0000-0003-4515-3468]

\authormark{KIVITS \textsc{et al.}}
\titlemark{Novel methodology for obtaining design structure matrices using network identification}

\address[1]{\orgname{DIFFER - Dutch Institute for Fundamental Energy Research, }%
\orgaddress{\street{De Zaale 20, }\postcode{5612 AJ, }\city{Eindhoven, }\country{the Netherlands}}}

\address[2]{\orgdiv{Department of Electrical Engineering, }\orgname{Eindhoven University of Technology, }%
\orgaddress{\postcode{5600 MB, }\city{Eindhoven, }\country{the Netherlands}}}

\address[3]{\orgdiv{Department of Mechanical Engineering, }\orgname{Eindhoven University of Technology, }%
\orgaddress{\postcode{5600 MB, }\city{Eindhoven, }\country{the Netherlands}}}

\corres{Lizan Kivits (\email{e.m.m.kivits@differ.nl})}



\fundingInfo{DIFFER is part of the institutes organization of NWO. This work has been funded by the Spherical Tokamak for Energy Production (STEP), a UKAEA programme to design and build a prototype fusion energy plant and a path to commercial fusion. This publication is part of the project Balls to the Wall (project no. 19695) of the research programme NWO Talent Programme VIDI, financed in part by the Dutch Research Council (NWO). This work has been carried out within the framework of the EUROfusion Consortium, funded by the European Union via the Euratom Research and Training Programme (Grant Agreement No 101052200 - EUROfusion).}


\transkeywords{design structure matrix | identification | network | topology | nuclear fusion}

\abstract[ABSTRACT]{Design structure matrices (DSMs) are used to comprehensively represent complex systems. They visualize and describe the dependencies between various variables, processes, states, and events. As such they are used in several system engineering approaches, such as requirement and interface management, fault detection, and supervisory control. Currently, a DSM is typically built from knowledge of experts. This may lead to an incomplete or imbalanced DSMs. For instance, elements and links might be missing or superfluous. In this article, we propose a novel method to acquire the DSM using state-of-the-art network identification methods. This demonstrates a proof-of-principle of identifying DSMs from data as an additional tool to the standard heuristic approach. In the future, we plan to embed DSMs in system design and supervisory controllers. We apply this technique to identify the DSM of a fusion reactor modelled by a five-chamber plasma model describing the transport in a tokamak.}

\transabstract[transABSTRACT]{Design structure matrices (DSMs) are used to comprehensively represent complex systems. They visualize and describe the dependencies between various variables, processes, states, and events. As such they are used in several system engineering approaches, such as requirement and interface management, fault detection, and supervisory control. Currently, a DSM is typically built from knowledge of experts. This may lead to an incomplete or imbalanced DSMs. For instance, elements and links might be missing or superfluous. In this article, we propose a novel method to acquire the DSM using state-of-the-art network identification methods. This demonstrates a proof-of-principle of identifying DSMs from data as an additional tool to the standard heuristic approach. In the future, we plan to embed DSMs in system design and supervisory controllers. We apply this technique to identify the DSM of a fusion reactor modelled by a five-chamber plasma model describing the transport in a tokamak.}




\copyright{This is an open access article under the terms of the \href{Creative Commons Attribution-NonCommercial}{Creative Commons Attribution-NonCommercial} License, which permits use, distribution and reproduction in any medium, provided the original~work~is~properly cited and is not used for commercial purposes.
\\[5pt]
  ©  2026 The author(s) submitted to \textit{System Engineering} published by Wiley Periodicals LLC.}


\maketitle


\section{Introduction} \label{sec:intro}
Design structure matrices (DSMs) are powerful tools to visualize, analyse, and manage complex systems \cite{Steward1981} and they provide a clear and compact representation of the structure of these systems. A DSM is a square matrix or table that captures the dependencies among elements composing the system \cite{Eppinger2012,Browning2016}. Analysing the DSM provides insight into the function, structure, and behaviour of the system \cite{Wilschut2017}, which has proven value for understanding, designing, and optimizing complex system architectures \cite{DSMweb2026}. The application domains of DSMs include engineering management, financial systems, health care organization, and social networks \cite{Eppinger2012,Browning2016,DSMweb2026}. 

Typically, a DSM is built through a heuristic approach by interviewing experts and combining their knowledge \cite{Eppinger2012, Beernaert2024b}. This requires a rather complete understanding of the function, structure, and behaviour of the system. The disadvantage of this approach is that it can be time-consuming and that the resulting DSM depends on the available knowledge of the experts and on the kind and number of experts that have been consulted. This may lead to missing relevant dependencies, superfluous irrelevant dependencies, and an imbalanced DSM. Managing a complex system based on an inaccurate DSM induces miscommunications and fallacious decisions, which result in undesired and even unsafe behaviour of the system. 

A DSM is equivalent to an adjacency matrix, which captures the interconnection structure of a network. In a dynamic network, signals are linked to each other through dynamic (i.e. time-dependent) relations \cite{Goncalves2007,VandenHof2013}. The interconnection structure between these signals can be identified from the data, without estimating the full dynamics of the network. This only requires measurement data of the signals of interest. In a Bayesian approach, the dynamic interconnections are modelled as Gaussian processes with additional configuration variables to optimize the algorithm (referred to as hyperparameters) \cite{Chiuso2012}. The tuning of the hyperparameters is automated in the Bayesian model selection method \cite{Shi2019}. Alternatively, the covariance-matching approach is free from hyperparameters and prior assumptions on the model parameters \cite{Venkitaraman2020}. All these methods consider (filtered) white noise as external influence only and do not incorporate known (or controllable) external perturbation signals. 

Our goal is to automate the construction of a DSM when time-dependent system data is available, to overcome the dependency on expert knowledge. Based on the one-to-one relationship between DSMs and adjacency matrices, we propose to identify a DSM from data by applying dynamic network identification methods that incorporate external perturbation signals. The advantage of this approach is that it merely depends on the system itself and that prior knowledge on signals can be incorporated. The physical quantities of interest have to be selected and measured, but no further knowledge on the structure or behaviour of the system is required. As the structure is ultimately dependent on the measurement set, the identified DSM will exactly reflect the structure that is relevant for managing the system. 

This methodology can, for example, be applied to organizations and social networks \cite{Eppinger2012,Browning2016}, where the DSM represents the relationships or information exchanges among units of an organisation or between individual people. Data that contains these interactions can be automatically collected from agendas and meeting schedules. Another application lies in the domain of complex products and engineering systems, such as vehicles, software, machinery, and built environments \cite{Eppinger2012,Browning2016}. Recently, DSMs are also used in nuclear fusion \cite{Beernaert2023, Beernaert2024a, Beernaert2024b, Aben2024} and can in principle also be used in similar systems, like electrolysers and molten salt reactors. Using simulations or experiments, the interface variables between components can be measured, from which the interconnection structure between the components (i.e. the DSM) can automatically be identified. 

In this article, we automate the construction of a DSM by identifying the DSM from simulation data. For this, we employ dynamic network identification techniques. To be precise, we extend the Bayesian model selection method \cite{Shi2019} by incorporating perturbation signals. To demonstrate the potential of this methodology, we identify, from simulation data, the DSM of the particle transport in a fusion reactor described by a five-chamber plasma model \cite{Figueiredo2025}. The results show that incorporating perturbation signals indeed improves the DSM and that it is possible to handle disturbances that process through the plasma. 

The remaining of this article is structured as follows. In Section~\ref{sec:dsm} and~\ref{sec:dynamicnetwork}, respectively, the DSM and the dynamic network are introduced. Our approach to obtain a DSM from data is explained in Section~\ref{sec:approach}. Section~\ref{sec:fusion} includes a background on nuclear fusion reactors, because a simulation example of a fusion reactor is included in Section~\ref{sec:example} to illustrate the potential of our method. Possible next steps in this research are discussed in Section~\ref{sec:future} and Section~\ref{sec:conclusion} concludes the article.

\section{Design structure matrix} \label{sec:dsm}
The DSM is a technique to model complex systems and an effective tool to manage the design of these systems \cite{Steward1981}. It is a square matrix or table that captures the dependencies among elements composing the system \cite{DSMweb2026, Eppinger2012}. This yields a concise visualization of the interactions in the system. Analysing the DSM leads to more insight into the system, which can be used to derive suggestions for improvement. A DSM can also serve as basis for synthesising controllers \cite{Wilschut2017}. 

\begin{figure}
    \centering
    \subfloat[\textbf{a.} A binary DSM of six elements.]{
        \includegraphics[trim={1cm 1cm 17cm 0},clip,width=.47\linewidth]{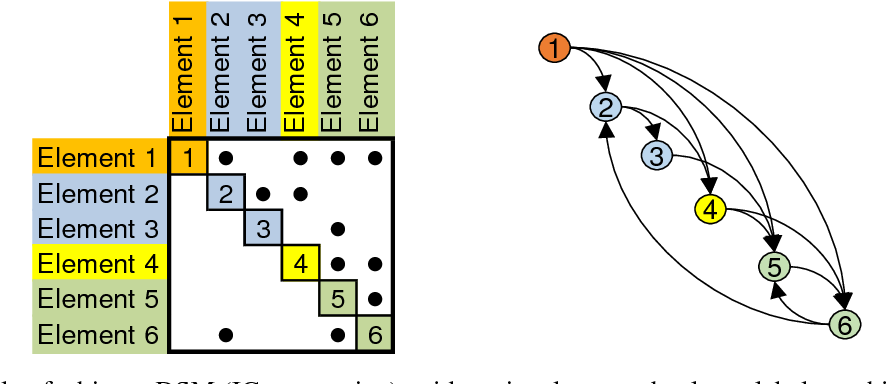}
        \label{fig:DSM_matrix}
    }\hfill
    \subfloat[\textbf{b.} A digraph of six nodes.]{
        \includegraphics[trim={18.5cm 1cm .5cm 1cm},clip,width=.46\linewidth]{DSM.png}
        \label{fig:DSM_graph}
    }
    \caption{A binary DSM (a) and its corresponding digraph (b) of a system consisting of six subsystems \cite{Browning2016}, where each subsystem is represented by an element in the DSM and a node in the digraph.}
    \label{fig:DSM}
\end{figure}

Figure~\ref{fig:DSM} shows an example of a DSM and its corresponding graphical visualisation as a directed graph (digraph) \cite{Browning2016}. They represent a system consisting of six subsystems, which are listed as elements in the DSM on the left-hand side and are visualized as nodes (circles) in the digraph on the right-hand side. The dependencies among the subsystems are represented by dots in the rows and columns of the corresponding elements in the DSM and by arrows linking the corresponding nodes in the digraph. 

The design structure of a system is also called the architecture of the system. The DSM can represent several architectures, such as organization structures (communication between people, teams, or departments), product architectures (interactions between components, subsystems, or functions), process architecture (information flow between activities or subprocesses), and engineering design processes (interactions between processes or parameters) \cite{Eppinger2012}. The DSM is typically built by interviewing experts and combining their knowledge. 

The DSM is analogous to the adjacency matrix in graph theory, which represents the adjacent edges between the vertices in a digraph \cite{Mesbahi2010}. The adjacency matrix of the digraph shown in Figure~\ref{fig:DSM_graph}, is given by: 
\begin{equation}\label{eq:Example_Adjacency}
    Adj = \begin{bmatrix}0&0&0&0&0&0\\ 1&0&0&0&0&1\\ 0&1&0&0&0&0\\ 1&1&0&0&0&0\\ 1&0&1&1&0&1\\ 1&0&0&1&1&0 \end{bmatrix}.
\end{equation}
This adjacency matrix is equivalent to the DSM in Figure~\ref{fig:DSM_matrix}, where each dot is replaced by a $1$, each empty box is replaced by a $0$, and the transpose of the matrix is taken. The transpose relation between the DSM and the adjacency matrix is only a matter of convention. Hence, the equivalence between an adjacency matrix and a DSM implies a one-to-one mapping between them.

The DSM is often binary, indicating whether there is a dependency (as the dot in Figure~\ref{fig:DSM_matrix}) or not. The strength of the dependency can also be quantified by a numerical value, analogous to the adjacency matrix of a weighted digraph. If the strength is quantified by a time-dependent system, such as a polynomial relation or a transfer function, the resulting matrix represents a dynamic graph or dynamic network\footnote{With a dynamic network, we mean a graph of which the edges are time-dependent subsystems. We do not mean a graph of which the structure evolves over time.}. As a result, there are three levels of representing the interconnections in a dynamic network: The dynamic relations can be described by a dynamic DSM, in which each entry is a mathematical expression; the strength of the dependencies can be described by a static DSM, in which each entry is a numerical value indicating the gain of the dependency; and the interconnection structure can be described by a binary DSM, in which each entry indicates the existence or absence of the dependency. 
The DSM is often binary, indicating whether there is a dependency (as the dot in Figure~\ref{fig:DSM_matrix}) or not. The strength of the dependency can also be quantified by a numerical value, analogous to the adjacency matrix of a weighted digraph. If the strength is quantified by a time-dependent system, such as a polynomial relation or a transfer function, the resulting matrix represents a dynamic graph or dynamic network\footnote{With a dynamic network, we mean a graph of which the edges are time-dependent subsystems. We do not mean a graph of which the structure evolves over time.}. As a result, there are three levels of representing the interconnections in a dynamic network: The dynamic relations can be described by a dynamic DSM, in which each entry is a mathematical expression; the strength of the dependencies can be described by a static DSM, in which each entry is a numerical value indicating the gain of the dependency; and the interconnection structure can be described by a binary DSM, in which each entry indicates the existence or absence of the dependency. 

\section{Dynamic network} \label{sec:dynamicnetwork}
A dynamic network is a system that consists of interconnected dynamic (i.e. time-dependent) subsystems. It can also be viewed as a set of signals that are linked to each other through dynamic relations \cite{Goncalves2007}. A dynamic network can be visualized by a digraph, in which the signals are the vertices and the dynamic subsystems or relations (so-called modules) are the edges. A dynamic network that is represented in terms of input-output modules is referred to as a module representation \cite{VandenHof2013}. A dynamic network in which the signals share information through their interconnections, rather than having directed (input-output) relationships, can be visualized by an undirected graph \cite{Willems2007} and is referred to as a diffusively coupled network \cite{Kivits2023}. 

A directed dynamic network consists of measured state signals $x_j(t)$, $j=1,2,\ldots,L$, that are interconnected through transfer function modules. Each state signal $x_j(t)$ can be described by a function of internal and external signals according to \cite{VandenHof2013}
\begin{equation} \label{eq:Network_state}
    x_j(t) = \sum_i G_{ji}(q) x_i(t) + \sum_k R_{jk}(q)u_k(t) + H_j(q) e_j(t),
\end{equation}
where $t$ indicates the time step; $G_{ji}(q)$, $R_{jk}(q)$, and $H_j(q)$ are stable transfer functions with $q^{-1}$ the delay operator meaning $q^{-1}w(t) = w(t-q)$; $v_j(t) := H_j(q) e_j(t)$ is the disturbance acting on $x_j(t)$, which is modelled as filtered white noise with monic minimum-phase filter $H_j(q)$ and white noise $e_j(t)$ with covariance $\sigma^2_j$; and $u_k(t)$, $k=1,2,\ldots,K$ is a known external perturbation signal acting on $x_j(t)$. The model \eqref{eq:Network_state} can be written in matrix form as 
\begin{equation}\label{eq:Network_matrix}
    x(t) = G(q)x(t) + R(q)u(t) + H(q)e(t),
\end{equation}
where $x(t)$ and $e(t)$ are row vectors of dimension $L\times 1$ with $j$th element $x_j(t)$ and $e_j(t)$, respectively; $u(t)$ is a row vector of dimension $K\times 1$ with $k$th element $u_k(t)$; $G(q)$ is a transfer function matrix having $(j,i)$th element $G_{ji}(q)$; $R(q)$ is a transfer function matrix having $(j,k)$th element $R_{jk}(q)$; $H(q)$ is a monic, stable, and stably invertible transfer function matrix with $(j,j)$th element $H_j(q)$. Observe that $G(q)$ has zeros on the diagonal and its zero-structure represents the interconnection structure between the states. It can be viewed as a dynamic adjacency matrix or dynamic DSM. 

\begin{figure}
    \centering
    \includegraphics[width=\linewidth]{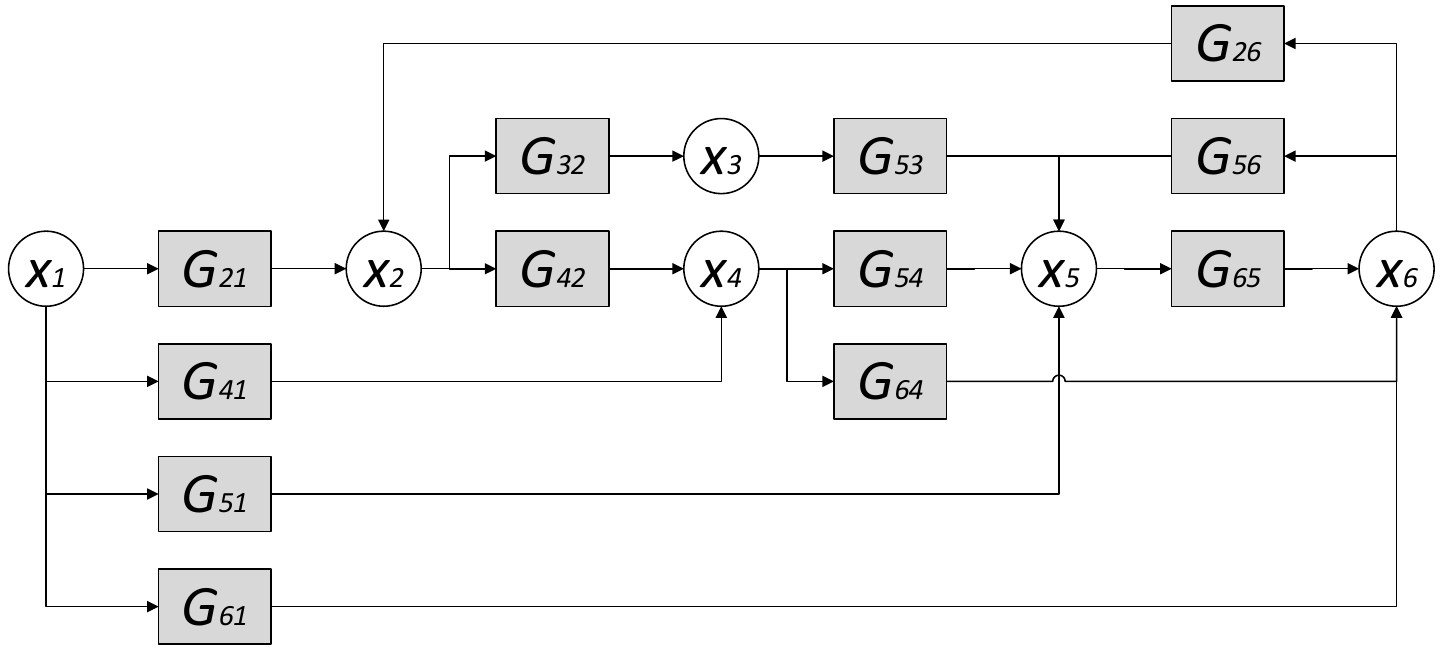}
    \caption{The module representation of the dynamic network corresponding to the digraph in Figure~\ref{fig:DSM_graph}.}
    \label{fig:Modrep}
\end{figure}

Figure~\ref{fig:Modrep} shows the module representation of the dynamic network corresponding to the digraph in Figure~\ref{fig:DSM_graph}. Its mathematical description is given by 
\begin{equation*}
    \resizebox{\linewidth}{!}{%
    $\underbrace{\begin{bmatrix}x_1(t)\\x_2(t)\\x_3(t)\\x_4(t)\\x_5(t)\\x_6(t)\end{bmatrix}}_{x(t)} = 
    \underbrace{\begin{bmatrix}0&0&0&0&0&0\\ G_{21}(q)&0&0&0&0&G_{26}(q)\\ 0&G_{32}(q)&0&0&0&0\\ G_{41}(q)&G_{42}(q)&0&0&0&0\\ G_{51}(q)&0&G_{53}(q)&G_{54}(q)&0&G_{56}(q)\\ G_{61}(q)&0&0&G_{64}(q)&G_{65}(q)&0\end{bmatrix}}_{G(q)} 
    \underbrace{\begin{bmatrix}x_1(t)\\x_2(t)\\x_3(t)\\x_4(t)\\x_5(t)\\x_6(t)\end{bmatrix}}_{x(t)}$,
    }
\end{equation*}
where $x(t)$ captures the six measured state signals of the system and $G(q)$ contains the time-dependent dynamic relations between them. The states $x_i(t)$, $i=1,\ldots,6$, are represented by the nodes in the digraph in Figure~\ref{fig:DSM_graph}. Observe that $G(q)$ has the same zero-structure as the adjacency matrix $Adj$ in \eqref{eq:Example_Adjacency} and thus can be viewed as a dynamic DSM. 

\section{DSM identification approach} \label{sec:approach}
The aim in topology identification for dynamic networks is to find the interconnection structure (topology) of the network from data. These methods aim to identify a binary matrix with the same zero-structure as $G(q)$. This adjacency matrix represents the topology of the graph and is equivalent to a DSM. Our idea is to use these topology identification methods to identify a DSM from data. 

Following the dynamic network identification community, we choose a Bayesian model selection approach \cite{Wasserman2000}, where the infinite impulse responses of the interconnections are modelled as Gaussian processes with hyperparameters \cite{Chiuso2012}. The hyperparameters can be estimated by maximizing the marginal likelihood through the expectation-maximization (EM) algorithm \cite{Bottegal2016}. We choose the Bayesian identification algorithm presented in \cite{Shi2019}, because prior information on the topology can be incorporated and tuning of the hyperparameters is automated through the EM algorithm. 

The best topology $\mathcal{G}$ is chosen to be the one that explains the measured data $z(t):=\{x(t),u(t)\}$ best. This is the topology with the highest marginal log-likelihood $\log P\left( z(t) | \mathcal{G} \right)$. This function can be decomposed into independent terms, each of which corresponding to the paths towards a specific state signal $x_j(t)$. Each independent term thus corresponds to all paths towards a specific state and therefore, the network can be decomposed into independent subparts, where each subpart includes all paths towards a specific state. The number of subparts is thus equal to the number of states and every subpart has multiple inputs and a single output. The topology of each independent multiple-input single-output (MISO) subpart of the network $\mathcal{G}_j$ can be formulated independently, based on data $z_j(t):=\{x(t),u_j(t)\}$, as:
\begin{equation}
    \max_{\mathcal{G}_j} \log P\left( z_j(t) | \mathcal{G}_j \right).
\end{equation}

Following the procedure in \cite{Shi2019}, we include the perturbation signals $u_k(t)$, $k=1,2,\ldots,K$, and the corresponding dynamics $R_{jk}(q)$ into the analysis. Using the joint-direct one-step-ahead predictor \cite{Ljung1999,Weerts2016}, the model \eqref{eq:Network_state} can be written as
\begin{multline} \label{eq:network_state_past}
    x_j(t) = (1-H_j^{-1}(q))x_j(t) + \sum_i H_j^{-1}(q)G_{ji}(q) x_i(t) \\
    + \sum_k H_j^{-1}(q) R_{jk}(q)u_k(t) + e_j(t).
\end{multline}
The finite ($n$th)-order approximation of $x_j(t)$ up to time $N$ is 
\begin{equation}\label{eq:network_state_N}
    x_j^N = A_j\theta_j + B_j\vartheta_{j} + e_j^N,
\end{equation}
where the vectors $x_j^N$ and $e_j^N$ contain the values of $x_j(\tau)$ and $e_j(\tau)$ for $\tau = 1,2,\ldots,N$, respectively; the matrix $A_j$ consists of Toeplitz matrices containing the measured state signals $x_i(\tau)$, $i=1,2,\ldots,L$, $\tau = 0,1,\ldots,N-1$; the matrix $B_j$ is a Toeplitz matrix containing the perturbation signal $u_k(\tau)$, $k=1,2,\ldots,K$, $\tau = 0,1,\ldots,N-1$; the vector $\theta_j$ contains the coefficients of the impulse responses from the states $x_i(t)$, $i=1,2,\ldots,L$, to state $x_j(t)$; and the vector $\vartheta_j$ contains the coefficients of the impulse response from $u_j(t)$ to $x_j(t)$. 

By following a kernel-based system identification approach, the impulse response coefficients $\theta_j$ and $\vartheta_j$ are modelled as zero-mean Gaussian processes with diagonal covariance matrix $K_j(\lambda_j)$ and $\kappa_j(\beta_j)$, respectively, with hyperparameter vectors $\lambda_j$ and $\beta_j$ that have only positive elements. We take the same structure for the covariance matrices as in \cite{Shi2019}. The cost function of the $j$th MISO problem becomes
\begin{align}
    J(\mathcal{G}_j;\eta_j) &= 2\log P\left( z_j^N|\mathcal{G}_j;\eta_j \right) - c, \label{eq:network_cost1} \\
                            &= -(x_j^N)^{\top}\Gamma_j^{-1}(x_j^N)^{\top} - \log\det(\Gamma_j), \label{eq:network_cost2}
\end{align}
where $\eta_j=\begin{bmatrix}\sigma_j&\lambda_j^{\top}&\beta_j^{\top}\end{bmatrix}^{\top}$ contains the hyperparameters; $c$ is a constant term that does not influence the optimum; and $\Gamma_j=\sigma_j^2 I_N + A_jK_jA_j^{\top} + B_j\kappa_jB_j^{\top}$ is the covariance matrix of $x_j^N$ resulting from modelling $\theta_j$, $\vartheta_j$, and $e_j^N$ as Gaussian processes in \eqref{eq:network_state_N}. 

The hyperparameters $\eta_j$ are iteratively estimated from maximizing the marginal log-likelihood \eqref{eq:network_cost2} for each MISO topology. Starting from an initial guess, the EM algorithm finds a local optimum \cite{Bottegal2016}. This value for the hyperparameters is then used in the marginal log-likelihood \eqref{eq:network_cost2} to find the best topology $\mathcal{G}_j$. 

As it is time-consuming to check for all possible topologies, a forward-backward greedy search algorithm has been implemented \cite{Shi2019}. The algorithm starts with an initial topology (e.g. an empty one) and iteratively adds the connection that improves the cost function the most until the cost function cannot be increased any more. Then connection are iteratively removed in a similar way to further improve the cost function.

The original topology identification algorithm \cite{Shi2019} has been implemented without external perturbation signals $u(t)$, that is with $R(q)=0$. We adapted the algorithm to incorporate $u(t)$ and $R(q)$ in several ways: 
\begin{enumerate}[{Case} 1.]\setcounter{enumi}{-1} 
    \item $R(q)=0$, which is the original algorithm \cite{Shi2019}.
    \item $R(q)=H(q)$, which is a generalisation of $R(q)=I_L=H(q)$, leading to $x_j^N = A_j\theta_j + u_j^N + e_j^N$ \eqref{eq:network_state_N}. 
    \item $R(q)$ is diagonal, that is $R_{jk}(q)=0$ for $j\neq k$. 
    \item $R(q)$ is free, that is $R_{jk}(q)$ is free for all $j,k$.
\end{enumerate}
For Case 0, the inputs $u_k(t)$ that are present are completely ignored in the identification procedure. For Case 1 and Case 2, it is assumed that it is known which inputs $u_k(t)$ enter at which states $x_j(t)$ and this structure is incorporated into the algorithm. These two cases are the most natural in practice, because it is typically known where the input signals enter the system. For Case 3, the interconnection structure from the inputs $u_k(t)$ towards the states $x_j(t)$ is assumed to be fully unknown and identified by the algorithm. 

\section{Fusion reactors} \label{sec:fusion}
The particular application of DSMs that we will focus on includes the plasma in fusion reactors. In magnetically confined fusion research \cite{Wesson2003}, a current carrying plasma is confined using magnetic fields to optimize the reactivity of deuterium and tritium. The overall tokamak control system includes the plasma control system (PCS), which deals with both the continuous and the discrete plasma behaviour \cite{Beernaert2024a}. An advanced supervisory control system needs to be incorporated to deal with machine constraints, hardware failure and plasma limits \cite{Clinque2020, DeVries2024, Beernaert2024a, Treutterer2017}. 

A key example of control in fusion reactors is the control of the heat and particle exhaust on the divertor target and the control of the plasma density. The exhaust can be controlled by impurity injection as well as the injection of deuterium gas \cite{Kool2025} to control the location of the impurity emission front \cite{Ravensbergen2021,Koenders2022,Koenders2023a,Koenders2023b}, the X-point radiator \cite{Bosman2025}, or the radiated power \cite{Ceelen2026}. The plasma density can be controlled in real-time by high-velocity introduction of DT ice, called pellet injection \cite{Derks2024}. The required feedback controllers are typically designed by a systematic model-based loop-shaping approach, which requires a model of the input-output dynamics, ideally including the plasma dynamics. Such model can be obtained from grey box modelling or can be identified from simulation data or experimental data e.g. via a frequency-domain approach \cite{vanBerkel2026}. Then, a locally linearized model is fit to the data and possible transient effects are removed by applying the local polynomial method (LPM) \cite{Schoukens2009,vanBerkel2020}. The resulting nonparametric model can directly be used for controller design, though a parametric model can also be estimated first. 

In contrast, the overall supervisory controller for the PCS can be synthesized based on the presence and absence of relations between the plasma variables, without including all the dynamics. A DSM can describe the relationships between actuators, plasma quantities, and sensors \cite{Beernaert2023,Beernaert2024b}. The elements of the physical aspects of the PCS are the finite plasma states and transitions, the continuous plasma processes and variables, and the requirements \cite{Beernaert2024a}. Actuators and sensors are used to interface with the PCS. The DSM contains the dependencies among all these elements. Analysing the DSM leads to a partitioning into several distinct clusters, each of which interacting with a subset of sensors and actuators. The clusters have strong internal interactions and minor interactions with the other clusters. For each cluster, a separate control strategy can be designed. Moreover, a DSM can be used to describe the architecture of the plasma that is managed by the PCS \cite{Aben2024} and as such, a DSM shows whether sufficient and\slash or redundant actuators and sensors are present for controlling the states of interest. Moreover, a DSM can serve as a basis for designing a supervisory controller for the PCS \cite{Aben2024}. 

The transitions between the different operating settings of the reactor can also be represented by a DSM. Depending on the operating condition of the reactor, different processes in the plasma are active or relevant. This can be modelled by discrete states, where each state represents a different operating condition. The possible transitions between the discrete states can be visualized in a DSM as well.

In this work, we will consider the ‘Spherical Tokamak for Energy Production’ (STEP), which aims at demonstrating the production of electricity with a nuclear fusion reactor and with this, promote the building of commercial nuclear fusion power plants \cite{Lennholm2024a}. The spherical design of the tokamak generally leads to higher energy production for a smaller volume, but also to more severe plasma control challenges. In addition, the plasma must be controlled using a limited amount of external actuators and sensors \cite{Lennholm2024b}, which further challenges the PCS design. 

Furthermore, STEP contains two divertors, one at the top (the upper divertor) and one at the bottom (the lower divertor) of the core. Both divertors have an inner and an outer leg, meaning that the tokamak can be represented by five chambers \cite{Figueiredo2025}: the core and the four divertor legs. In addition, both outer divertor legs are connected to a pump that disposes particles from the tokamak. Chamber models describing the particle transport are an accepted, albeit simplistic, approach for control-oriented modelling of tokamaks \cite{Derks2026, Figueiredo2025}.

\begin{figure}
    \begin{minipage}[b]{0.49\linewidth}
        \centering
        \includegraphics[width=\linewidth]{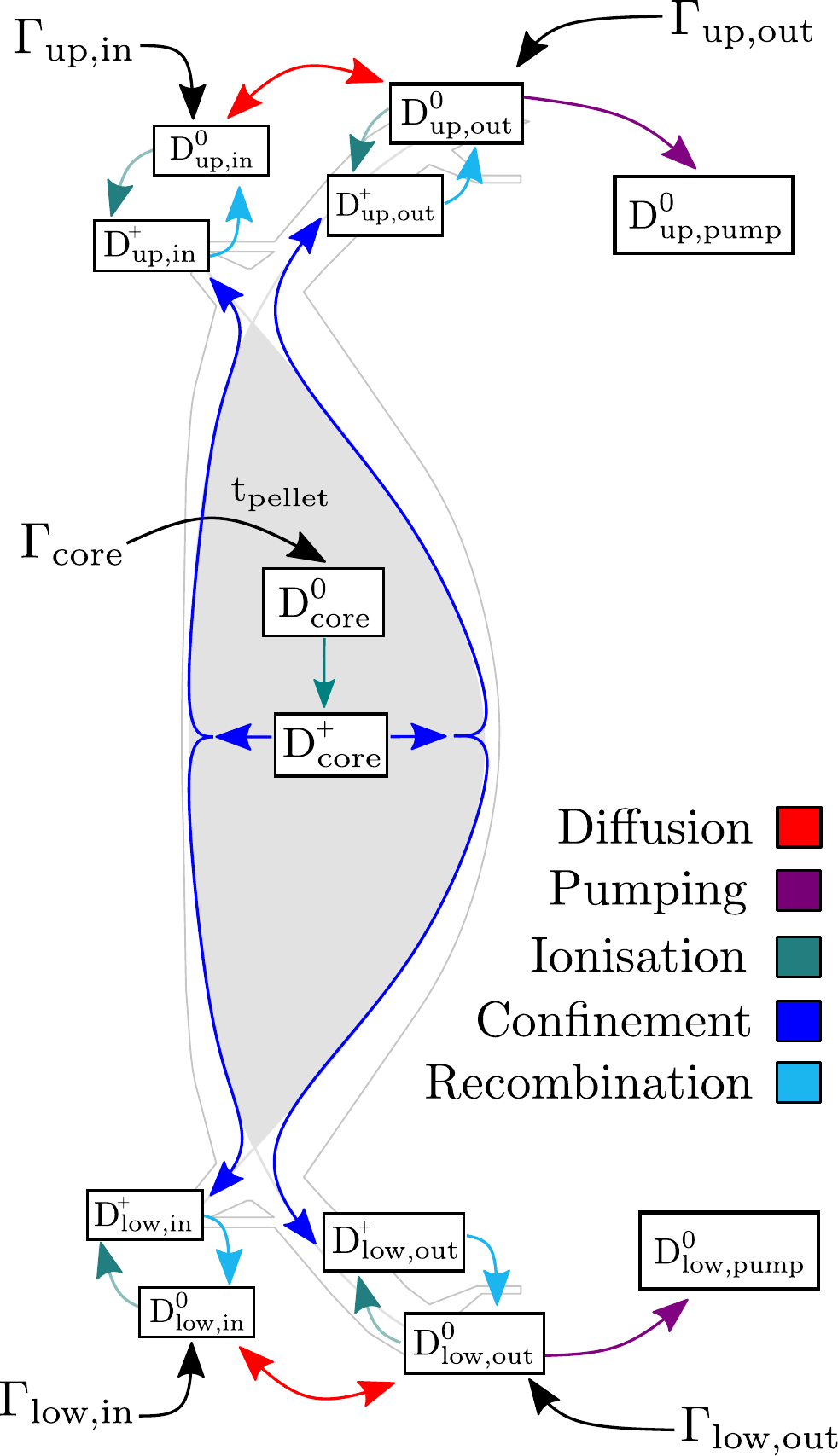}
        \caption{The five-chamber tokamak with a core and an upper and a lower divertor \cite{Figueiredo2025}. $\Gamma_a$ represents the gas inlet flow, consisting of neutral deuterium particles, in divertor leg $a$. $D^0_b$ and $D^+_b$ represent the number of neutrals (neutral deuterium particles) and the number of plasma particles (ionized deuterium particles), respectively, at location $b$.}
        \label{fig:Tokamak}
    \end{minipage}\hfill
    \begin{minipage}[b]{0.49\linewidth}
        \centering
        \resizebox{\linewidth}{!}{\begin{tikzpicture}[
    myneut/.style={circle,draw=black, fill=black!10, minimum size=20pt, inner sep=0pt, font=\footnotesize},
    myplas/.style={circle,draw=black, fill=black!30, minimum size=20pt, inner sep=0pt, font=\footnotesize},
    mypump/.style={circle,draw=black, fill=black!10, minimum size=20pt, inner sep=0pt, font=\footnotesize},
    mygas/.style={ circle,draw=black, fill=black!0,  minimum size=20pt, inner sep=0pt, font=\footnotesize},
    node distance=.5cm and .5cm, every text node part/.style={align=center}
    ]

    \node[myplas] (plas_c) {$x_{10}$};
    \node[myneut,left=1.5cm of plas_c] (neut_c) {$x_5$};
    \node[mygas, left=of neut_c] (gas_c) {$u_5$};   
   
    \node[myplas,above left=of plas_c] (plas_iu) {$x_7$};
    \node[myneut,above=of plas_iu] (neut_iu) {$x_2$};
    \node[mygas, left=of neut_iu] (gas_iu) {$u_2$};

    \node[myplas,above right=of plas_c] (plas_ou) {$x_6$};
    \node[myneut,above=of plas_ou] (neut_ou) {$x_1$};
    \node[mypump,right=of plas_ou] (pump_u) {$x_{11}$};
    \node[mygas, right=of neut_ou] (gas_ou) {$u_1$};
      
    \node[myplas,below left=of plas_c] (plas_il) {$x_9$};
    \node[myneut,below=of plas_il] (neut_il) {$x_4$};
    \node[mygas, left=of neut_il] (gas_il) {$u_4$};

    \node[myplas,below right=of plas_c] (plas_ol) {$x_8$};
    \node[myneut,below=of plas_ol] (neut_ol) {$x_3$};
    \node[mypump,right=of plas_ol] (pump_l) {$x_{12}$};
    \node[mygas, right=of neut_ol ] (gas_ol) {$u_3$};

    \foreach \i/\j in {
      gas_ou/neut_ou,   gas_iu/neut_iu, 
      gas_ol/neut_ol,   gas_il/neut_il,
      gas_c/neut_c,     neut_c/plas_c,
      neut_ou/pump_u,   neut_ol/pump_l}
      \path[->, >={Latex[length=2mm]}]
      (\i) edge (\j);
    
    \foreach \i/\j in {
      neut_ou/neut_iu,  neut_iu/neut_ou,
      neut_ol/neut_il,  neut_il/neut_ol,
      neut_ou/plas_ou,  plas_ou/neut_ou,
      neut_ol/plas_ol,  plas_ol/neut_ol,
      neut_iu/plas_iu,  plas_iu/neut_iu,
      neut_il/plas_il,  plas_il/neut_il}
      \path[->, >={Latex[length=2mm]}]
      (\i) edge[bend left=10] (\j);

    \foreach \i/\j in {plas_c/plas_ol,plas_c/plas_iu}
      \path[->,>={Latex[length=2mm]}, dashed]
      (\i) edge[bend left=20] (\j);

    \foreach \i/\j in {plas_c/plas_ou,plas_c/plas_il}
      \path[->,>={Latex[length=2mm]}, dashed]
      (\i) edge[bend right=20] (\j);
      
\end{tikzpicture}}
        \vspace{1.2cm}
        \caption{Digraph of the five-chamber model (without self-loops), showing the vertices (circles) with the gas inlets in the five chambers (white), the number of plasma particles in the five chambers (dark grey), and the number of neutrals in the five chambers and the pumps (light gray). The edges (arrows) show the particle flows, where the dotted edges indicate a relative small particle flow.}
        \label{fig:Five_chamber_digraph}
    \end{minipage}
\end{figure}

\section{Simulation example} \label{sec:example}

\subsection{Five-chamber model}
To illustrate the potential of our approach described in Section~\ref{sec:approach}, we apply this method to a five-chamber model of a tokamak, like STEP. The particle transport in such a tokamak can be modelled with the five-chamber model \cite{Figueiredo2025} as shown in Figure~\ref{fig:Tokamak}. This model describes the number of ions (ionized deuterium particles) and the number of neutrals (neutral deuterium particles) in the five chambers: the core and the four divertor legs. It also includes the number of neutral particles in the pumps, which are responsible for removing particles from the outer divertor legs. These 12 quantities are the (measured) states of the state-space model of the tokamak and are given in Table~\ref{tab:States}. Neutral deuterium particles are added to all five chambers through gas inlets (either continuous flows or discrete pellet injections), being the five input signals $u_i$, $i=1,2,\ldots,5$, of the state-space model. Every state $x_j$ is subject to a process disturbance signal $v_j(t)$. The resulting state-space model is 
\begin{equation}\label{eq:statespace}
    \dot{x}(t)=Ax(t)+Bu(t)+v(t), \qquad y(t)=x(t),
\end{equation}
with state matrix 
\begin{equation*}
    \resizebox{\linewidth}{!}{%
    $A=\begin{bmatrix}
        -1093&  2933&     0&     0&     0&  263&     0&    0&     0&      0& 0& 0 \\
           92& -4933&     0&     0&     0&    0&  1000&    0&     0&      0& 0& 0 \\
            0&     0& -1093&  2933&     0&    0&     0&  263&	  0&      0& 0& 0 \\
            0&     0&    92& -4933&     0&    0&     0&    0&  1000&      0& 0& 0 \\
            0&     0&     0&     0& -2000&    0&     0&    0&     0&      0& 0& 0 \\
         1000&     0&     0&     0&     0& -263&     0&    0&	 0&  0.042& 0& 0 \\
            0&  2000&     0&     0&     0&    0& -1000&    0&     0&  0.008& 0& 0 \\
            0&     0&  1000&     0&     0&    0&     0& -263&     0&  0.042& 0& 0 \\
            0&     0&     0&  2000&     0&    0&     0&    0& -1000&  0.008& 0& 0 \\
            0&     0&     0&     0&  2000&    0&     0&    0&     0& -0.101& 0& 0 \\
        1.407&     0&     0&     0&     0&    0&     0&    0&     0&      0& 0& 0 \\
            0&     0& 1.407&     0&     0&    0&     0&    0&     0&      0& 0& 0
    \end{bmatrix}$%
    }
\end{equation*}
and input matrix $B = \begin{bmatrix}I_5& 0_{5,7}\end{bmatrix}^{\top}$, with $I_5$ the identity matrix of size $5\times5 $ and $0_{5,7}$ a zero matrix of size $5\times 7$. The state-space model \eqref{eq:statespace} can be converted into a module representation \eqref{eq:Network_matrix} similar to as described in \cite{Kivits2018}. Moreover, the state matrix $A$ can directly be converted into an adjacency matrix or binary DSM by setting all non-zero elements to $1$. This is exactly the DSM that we aim to identify from the simulation data. The corresponding digraph of the five-chamber tokamak model is shown in Figure~\ref{fig:Five_chamber_digraph}, where the vertices represent inputs and states and the edges represent causal influences. The dotted edges indicate a weak relationship. For simplification, self-loops are omitted in the graph. From Figure~\ref{fig:Five_chamber_digraph}, it can be seen that the true graph $\mathcal{G}_0$ has 19 edges between the states and 125 non-existing edges between the states. 

\begin{table}[tb]
    \centering
    \caption{The physical meaning of the 12 states of the tokamak model, which is depicted in a digraph in Figure~\ref{fig:Five_chamber_digraph}. }
    \label{tab:States}
    \resizebox{\linewidth}{!}{%
    \begin{tabular}{l|l|l}
        State        & Measure                         & Location \\ \hline
        \lgc$x_1$    & \lgc Number of neutrals         & \lgc Upper divertor, outer leg \\
        \lgc$x_2$    & \lgc Number of neutrals         & \lgc Upper divertor, inner leg \\
        \lgc$x_3$    & \lgc Number of neutrals         & \lgc Lower divertor, outer leg \\
        \lgc$x_4$    & \lgc Number of neutrals         & \lgc Lower divertor, inner leg \\
        \lgc$x_5$    & \lgc Number of neutrals         & \lgc Core \\
        \dgc$x_6$    & \dgc Number of plasma particles & \dgc Upper divertor, outer leg \\
        \dgc$x_7$    & \dgc Number of plasma particles & \dgc Upper divertor, inner leg \\
        \dgc$x_8$    & \dgc Number of plasma particles & \dgc Lower divertor, outer leg \\
        \dgc$x_9$    & \dgc Number of plasma particles & \dgc Lower divertor, inner leg \\
        \dgc$x_{10}$ & \dgc Number of plasma particles & \dgc Core \\
        \lgc$x_{11}$ & \lgc Number of neutrals         & \lgc Upper pump \\  
        \lgc$x_{12}$ & \lgc Number of neutrals         & \lgc Lower pump
    \end{tabular}}
\end{table}

\subsection{Setup} \label{subsec:setup}
We perform two sets of simulations, where each set consists of three simulations: one with only input signals ($v(t)=0$); one with only disturbance signals ($u(t)=0$); and one with both input signals and disturbance signals acting on the network. The difference between the two sets of simulations lies in the amplitude of the disturbances. For all simulations, the five gas inlets are pseudorandom binary sequences (prbs) signals with amplitudes as given in Table~\ref{tab:Signals}. The disturbances $v_j$ are uniformly distributed random values in the interval $[-a,a]$, with $a$ the amplitude given in Table~\ref{tab:Signals} for both sets of simulations. As the disturbances process through the network, they are different from measurement noise. All states are measured over a time period of 100 seconds with a sampling time of 1 millisecond, leading to $N=100.001$ samples. 

\begin{table}[tb]
    \centering
    \caption{The amplitudes of the input signals $u_j$, $j=1,2,\ldots,5$, and the amplitudes of the disturbances $v_i$ entering at the states $x_i$, $i=1,2,\ldots,12$, for simulations set 1 and 2. }
    \label{tab:Signals}
    \resizebox{\linewidth}{!}{%
    \begin{tabular}{c||c|c|c|c|c}
        Input       &  $u_1$ & $u_2$ & $u_3$ & $u_4$ & $u_5$\\ \hline
        Amplitude   &  $1\times10^{21}$ & $3\times10^{20}$ & $1\times10^{21}$ & $3\times10^{20}$ & $5\times10^{20}$\\ \hline\hline
        Disturbance &  $v_1$ & $v_2$ & $v_3$ & $v_4$ & $v_5$\\ \hline
        Amplitude 1 &  $1\times10^{19}$ & $3  \times10^{17}$ & $1\times10^{19}$ & $3  \times10^{17}$ & $5  \times10^{16}$\\ \hline
        Amplitude 2 &  $5\times10^{20}$ & $1.5\times10^{20}$ & $5\times10^{20}$ & $1.5\times10^{20}$ & $2.5\times10^{20}$\\ \hline\hline
        Disturbance &  $v_6$ & $v_7$ & $v_8$ & $v_9$ & $v_{10}$\\ \hline
        Amplitude 1 &  $5\times10^{18}$ & $1.5\times10^{17}$ & $5\times10^{18}$ & $1.5\times10^{17}$ & $2.5\times10^{19}$ \\ \hline
        Amplitude 2 &  $5\times10^{20}$ & $1.5\times10^{20}$ & $5\times10^{20}$ & $1.5\times10^{20}$ & $2.5\times10^{20}$ \\ \hline\hline
        Disturbance &  $v_{11}$ & $v_{12}$ & ~ & ~ & ~\\ \hline
        Amplitude 1 &  $1\times10^{21}$ & $1\times10^{21}$ & ~ & ~ & ~ \\ \hline
        Amplitude 2 &  $5\times10^{20}$ & $5\times10^{20}$ & ~ & ~ & ~ \\ 
    \end{tabular}}
\end{table}

For identification purposes, the measurement data is scaled with a factor $10^{-21}$ (for numerical stability) and down-sampled with a sample period of 100 samples leading to a data length of $N=1001$ samples (for time reduction). The initial values of the hyperparameters are set in line with \cite{Shi2019} to $\sigma = 0.1$, $\beta = 0.5$, and $\lambda = 0.5$. The stopping criterium for the EM algorithm determining the hyperparameters is set to $\gamma = 0.2$. The order of the finite impulse response estimates is chosen to be $n=100$ as in \cite{Shi2019}. 

To evaluate the quality of the results, the true positive rate (TPR) and the false positive rate (FPR) are quantified. The TPR is the ratio between the number of existing edges that is correctly estimated and the total number of existing edges in the true graph. Its ideal value is 1. The FPR is the ratio between the number of non-existing edges that is incorrectly estimated (i.e. superfluously estimated edges) and the total number of non-existing edges in the true graph. Its ideal value is 0. The TPR and FPR are used to determine the distance ($dist$) from the estimated topology to the true one as
\begin{equation}\label{eq:distance}
    dist = \sqrt{FPR^2+(1-TPR)^2}.
\end{equation}

\begin{figure}[htbp]
    \centering
    \subfloat[\textbf{a.} TPR and FPR of the estimated topologies for the network subject to relatively \emph{small} disturbances]{
        \includegraphics[width=.46\linewidth]{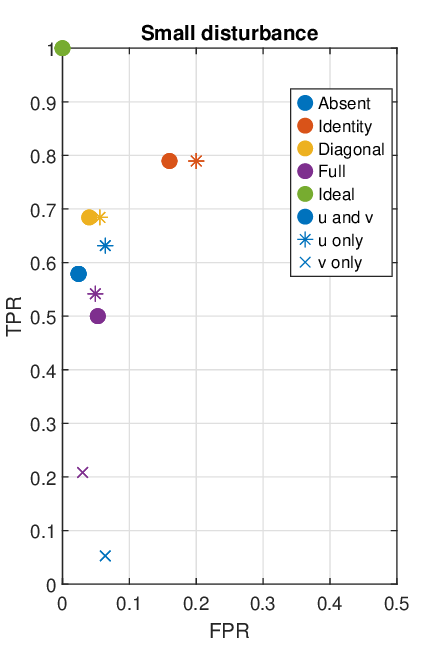}
        \label{fig:Distance_a}
    }
    \hfill
    \subfloat[\textbf{b.} TPR and FPR of the estimated topologies for the network subject to relatively \emph{large} disturbances]{
        \includegraphics[width=.46\linewidth]{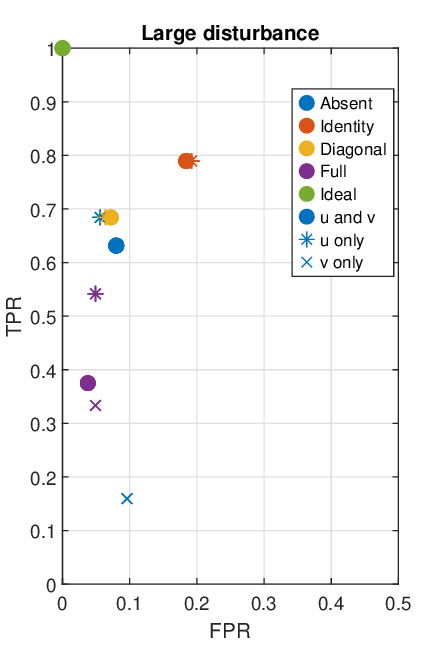}
        \label{fig:Distance_b}
    }
    \caption{TPR and FPR of the estimated topologies with different incorporation of the input term $R(q)u(t)$ in the algorithm: $R(q)=0$ (Case 0, blue), $R(q)=H(q)=I$ (Case 1, orange), $R(q)$ is diagonal (Case 2, yellow), $R(q)$ is free (Case 3, purple); and under different perturbation conditions: both inputs $u(t)$ and disturbances $v(t)$ ($\bullet$); only inputs $u(t)$ ($\ast$); or only disturbances $v(t)$ ($\times$). The ideal result is indicated in green. }
    \label{fig:Distance}
\end{figure}
\subsection{Results}
Figure~\ref{fig:Distance} shows the TPR and FPR for all simulations described in Section~\ref{subsec:setup} and all cases of the implemented algorithm, which are described in Section~\ref{sec:approach}. Case 1 and 2 are not shown for the simulations with only disturbances, because these algorithms do not apply to these simulations (and they give the same result as Case 0). The simulations with only disturbances give the worst results (largest distances) for all cases, especially, when the amplitude is small. For Case 1 and 2, the simulations with both inputs and disturbances give better results (smaller distances) than with only inputs. 

Figure~\ref{fig:Distance_a} and Figure~\ref{fig:Distance_b} show the TPR and FPR for the simulation set relative small and large amplitudes of the disturbances, respectively. From comparing Figure~\ref{fig:Distance_a} and Figure~\ref{fig:Distance_b}, we observe that for both simulation sets, Case 1 gives the smallest distance \eqref{eq:distance}, followed by Case 2, then Case 0, and finally, Case 3. To be precise, the best results (smallest distances) are obtained for the simulation with both inputs and disturbances and with the algorithm of Case 1, leading to a distance of $dist=0.2644$ and $dist=0.2796$ for the small and large disturbances, respectively. The identified adjacency matrices for these simulations are, respectively, 
\begin{equation*}
    \resizebox{\linewidth}{!}{%
    ${Adj}_{1}=\begin{bmatrix}
        \greone&  \greone&  \redzer&  \redzer&  \redzer&  \greone&  \redzer&  \redzer&  \redzer&  \grezer& \grezer& \grezer \\
        \greone&  \greone&  \redzer&  \grezer&  \redzer&  \redzer&  \greone&  \redzer&  \grezer&  \grezer& \grezer& \grezer \\
        \grezer&  \grezer&  \greone&  \greone&  \grezer&  \grezer&  \redzer&  \greone&  \redzer&  \redzer& \grezer& \grezer \\
        \grezer&  \redzer&  \greone&  \greone&  \grezer&  \grezer&  \grezer&  \redzer&  \greone&  \grezer& \grezer& \grezer \\
        \grezer&  \grezer&  \grezer&  \redzer&  \greone&  \grezer&  \redzer&  \grezer&  \grezer&  \redzer& \grezer& \grezer \\
        \greone&  \grezer&  \grezer&  \grezer&  \grezer&  \greone&  \grezer&  \grezer&  \redzer&  \redone& \grezer& \grezer \\
        \grezer&  \greone&  \grezer&  \grezer&  \grezer&  \grezer&  \greone&  \grezer&  \grezer&  \redone& \grezer& \grezer \\
        \grezer&  \grezer&  \greone&  \grezer&  \grezer&  \grezer&  \grezer&  \greone&  \grezer&  \redone& \grezer& \grezer \\
        \grezer&  \grezer&  \grezer&  \greone&  \grezer&  \grezer&  \grezer&  \grezer&  \greone&  \redone& \grezer& \grezer \\
        \grezer&  \grezer&  \grezer&  \grezer&  \greone&  \grezer&  \grezer&  \grezer&  \grezer&  \greone& \grezer& \grezer \\
        \greone&  \grezer&  \grezer&  \grezer&  \grezer&  \grezer&  \grezer&  \grezer&  \grezer&  \grezer& \grezer& \grezer \\
        \grezer&  \grezer&  \greone&  \grezer&  \redzer&  \grezer&  \grezer&  \grezer&  \grezer&  \grezer& \grezer& \grezer
    \end{bmatrix},\
    {Adj}_{2}=\begin{bmatrix}
         \greone&  \greone&  \grezer&  \redzer&  \grezer&  \greone&  \redzer&  \grezer&  \grezer&  \grezer& \grezer&  \redzer \\
         \greone&  \greone&  \redzer&  \redzer&  \grezer&  \redzer&  \greone&  \grezer&  \redzer&  \grezer& \grezer&  \grezer \\
         \grezer&  \grezer&  \greone&  \greone&  \redzer&  \grezer&  \grezer&  \greone&  \redzer&  \grezer& \grezer&  \grezer \\
         \redzer&  \grezer&  \greone&  \greone&  \redzer&  \grezer&  \grezer&  \redzer&  \greone&  \redzer& \grezer&  \grezer \\
         \grezer&  \redzer&  \grezer&  \grezer&  \greone&  \grezer&  \redzer&  \grezer&  \grezer&  \redzer& \grezer&  \grezer \\
         \greone&  \redzer&  \grezer&  \grezer&  \redzer&  \greone&  \redzer&  \grezer&	\grezer&  \redone& \grezer&  \grezer \\
         \grezer&  \greone&  \grezer&  \grezer&  \grezer&  \grezer&  \greone&  \grezer&  \grezer&  \redone& \grezer&  \grezer \\
         \grezer&  \redzer&  \greone&  \redzer&  \grezer&  \grezer&  \grezer&  \greone&  \redzer&  \redone& \grezer&  \grezer \\
         \grezer&  \grezer&  \grezer&  \greone&  \grezer&  \grezer&  \grezer&  \grezer&  \greone&  \redone& \grezer&  \grezer \\
         \grezer&  \redzer&  \grezer&  \grezer&  \greone&  \grezer&  \grezer&  \grezer&  \grezer&  \greone& \grezer&  \grezer \\
         \greone&  \grezer&  \grezer&  \grezer&  \grezer&  \grezer&  \grezer&  \grezer&  \grezer&  \grezer& \grezer&  \grezer \\
         \grezer&  \grezer&  \greone&  \grezer&  \grezer&  \grezer&  \grezer&  \grezer&  \grezer&  \grezer& \grezer&  \grezer
    \end{bmatrix},$%
    }
\end{equation*}
where the green zeros and ones indicate correctly identified links and where the red zeros and ones indicate unidentified and superfluously identified links, respectively. The fact that both cases result in a TPR of $0.7895$ is due to the fact that the four weak connections (the dotted arrows in Figure~\ref{fig:Five_chamber_digraph}) are not identified (as indicated in the adjacency matrices as red zeros). 

The adjacency matrices $Adj_1$ and $Adj_2$ are translated into DSMs given in Table~\ref{tab:DSM1} and Table~\ref{tab:DSM2}, respectively. The DSMs are obtained by replacing all ones in the adjacency matrices by dots, replacing all zeros by empty cells, and transposing the whole. This means that in the DSMs, bullets indicate links from the elements indicated at the left-hand side to the elements indicated at the top. Green bullets indicate correctly identified links and red bullets indicate superfluously identified links. Spaces indicate (correctly identified) non-existing links and red bars indicate missing links, i.e. links that are present in the actual underlying system, but that not have been identified. The grey colours in the DSMs match the states in Figure~\ref{fig:Five_chamber_digraph} and Table~\ref{tab:States}. 

From the DSMs in Table~\ref{tab:DSM1} and~\ref{tab:DSM2}, it follows that only the four weak connections (the dotted arrows in Figure~\ref{fig:Five_chamber_digraph}) are not identified, while all other links that are present in the underlying system are indeed identified. The fact that the weak connections are missed by the algorithms is most likely caused by their relative small gain. From the DSMs, it is clear that some more links have been identified than present in the underlying system. These superfluous links, represented by the red dots, are expected to correct for the dynamics that are actually caused by the four weak links. From the DSMs, it can be observed that these superfluous links are located closely to the correctly identified links. 

\begin{table}[tb]
    \centering
    \caption{The DSM corresponding to the adjacency matrix $Adj_1$.}
    \label{tab:DSM1}
    \resizebox{\linewidth}{!}{%
    \begin{tabular}{c||c|c|c|c|c|c|c|c|c|c|c|c|}
                ~& \lgc\drts{Number of neutrals}{Upper divertor, outer leg}& \lgc\drts{Number of neutrals}{Upper divertor, inner leg}
                 & \lgc\drts{Number of neutrals}{Lower divertor, outer leg}& \lgc\drts{Number of neutrals}{Lower divertor, inner leg}
                 & \lgc\drts{Number of neutrals}{Core}
                 & \dgc\drts{Number of plasma particles}{Upper divertor, outer leg}& \dgc\drts{Number of plasma particles}{Upper divertor, inner leg}
                 & \dgc\drts{Number of plasma particles}{Lower divertor, outer leg}& \dgc\drts{Number of plasma particles}{Lower divertor, inner leg} 
                 & \dgc\drts{Number of plasma particles}{Core}
                 & \lgc\drts{Number of neutrals}{Upper pump}& \lgc\drts{Number of neutrals}{Lower pump}\\ \hline \hline
        \lgc\drt{Number of neutrals}{Upper divertor, outer leg}         &\lgc\grebul&\grebul&   ~   &   ~   &   ~   &\grebul&   ~   &   ~   &   ~   &   ~   &\grebul&   ~    \\ \hline
        \lgc\drt{Number of neutrals}{Upper divertor, inner leg}         &\grebul&\lgc\grebul&   ~   &\redbul&   ~   &   ~   &\grebul&   ~   &   ~   &   ~   &   ~   &   ~    \\ \hline
        \lgc\drt{Number of neutrals}{Lower divertor, outer leg}         &\redbul&\redbul&\lgc\grebul&\grebul&   ~   &   ~   &   ~   &\grebul&   ~   &   ~   &   ~   &\grebul \\ \hline
        \lgc\drt{Number of neutrals}{Lower divertor, inner leg}         &\redbul&   ~   &\grebul&\lgc\grebul&\redbul&   ~   &   ~   &   ~   &\grebul&   ~   &   ~   &   ~    \\ \hline
        \lgc\drt{Number of neutrals}{Core}                              &\redbul&\redbul&   ~   &   ~   &\lgc\grebul&   ~   &   ~   &   ~   &   ~   &\grebul&   ~   &\redbul \\ \hline
        \dgc\drt{Number of plasma particles}{Upper divertor, outer leg} &\grebul&\redbul&   ~   &   ~   &   ~   &\dgc\grebul&   ~   &   ~   &   ~   &   ~   &   ~   &   ~    \\ \hline
        \dgc\drt{Number of plasma particles}{Upper divertor, inner leg} &\redbul&\grebul&\redbul&   ~   &\redbul&   ~   &\dgc\grebul&   ~   &   ~   &   ~   &   ~   &   ~    \\ \hline
        \dgc\drt{Number of plasma particles}{Lower divertor, outer leg} &\redbul&\redbul&\grebul&\redbul&   ~   &   ~   &   ~   &\dgc\grebul&   ~   &   ~   &   ~   &   ~    \\ \hline
        \dgc\drt{Number of plasma particles}{Lower divertor, inner leg} &\redbul&   ~   &\redbul&\grebul&   ~   &\redbul&   ~   &   ~   &\dgc\grebul&   ~   &   ~   &   ~    \\ \hline
        \dgc\drt{Number of plasma particles}{Core}                      &   ~   &   ~   &\redbul&   ~   &\redbul&\redbar&\redbar&\redbar&\redbar&\dgc\grebul&   ~   &   ~    \\ \hline
        \lgc\drt{Number of neutrals}{Upper pump}                        &   ~   &   ~   &   ~   &   ~   &   ~   &   ~   &   ~   &   ~   &   ~   &   ~   &\lgc   ~   &   ~    \\ \hline
        \lgc\drt{Number of neutrals}{Lower pump}                        &   ~   &   ~   &   ~   &   ~   &   ~   &   ~   &   ~   &   ~   &   ~   &   ~   &   ~   &\lgc   ~    \\ \hline
    \end{tabular}}
\end{table}

\subsection{Discussion}
With the original algorithm (shown in blue in Figure~\ref{fig:Distance}), a TPR between 0.5 and 0.7 is achieved when the input signals are present and the TPR is less than 0.2 when only disturbance signals are present. This was not expected, because in the literature \cite{Shi2019,VanEsch2020} it is possible to achieve a TPR of 0.9. We believe that the limited performance is caused by the sparsity of the network (where one missing link has a higher impact on the TPR) and the relative large differences between the gains of the interconnections (the networks in the literature \cite{Shi2019,VanEsch2020} have relative equal gains). The reason why the result improves (i.e. the distance decreases) for adding input signals is that the input signal has a random behaviour as well and therefore adds information to the data, even though the input signals are not incorporating in the algorithm. In addition, we know from experience that filtered white noise (coloured noise) disturbances lead to better results than unfiltered white noise disturbances. 

For Case 1, the FPR is around 0.2 and the TPR is around 0.8, which is an acceptable result, although there is some room for improvement. For the other algorithms, the FPR is smaller than 0.1, which is good, and the TPR is largely spread between 0.3 and 0.7, which is poor to moderate. We believe that the relative low TPR is caused by the sparsity of the network, because a superfluous link in the identified network has les impact than a missing link. Case 1 achieves a better balance between the FPR and the TPR, due to the most suitable incorporation of the input signals. 

Incorporating perturbation signals improves the result if the interconnection structure of the inputs is fixed. Providing the exact signals entering the network to the algorithm helps in identifying the interconnections, leading to less missing links and thus a higher TPR. Most links can be identified, with the exception of links with a small gain. This means that some dynamics in the network are not explained, because of the missing links. The missing links lead to a TPR unequal to 1. The algorithm is compensating this by including additional links that cause an FPR unequal to 0. This is even the situation when the input signals are incorporated in the algorithm, because the noise has a smaller covariance, as the information that comes from the perturbation signals is separated from the noise. As a result, the unexplained dynamics in the network are less likely to be explained by the noise and therefore, more superfluous interconnections are added to explain these dynamics. 

If the topology from the inputs to the states is completely left free, there is too much freedom in the structure of $R(q)$, causing too much options for directly connecting inputs to states. As more inputs to a state are possible, the MISO problem that has to be solved is larger (has more inputs), while the data set remains of the same size. This also degrades the results. 

Overall, the simulation example shows that it is possible to identify the links between the states from data, giving a DSM. Therefore, it is a promising approach to identify a DSM from data only. Improvements can be made in the future by improving the algorithm itself, by exploring other algorithms and by reconsidering the decision making. 
\begin{table}[tb]
    \centering
    \caption{The DSM corresponding to the adjacency matrix $Adj_2$.}
    \label{tab:DSM2}
    \resizebox{\linewidth}{!}{%
    \begin{tabular}{c||c|c|c|c|c|c|c|c|c|c|c|c|}
                ~& \lgc\drts{Number of neutrals}{Upper divertor, outer leg}& \lgc\drts{Number of neutrals}{Upper divertor, inner leg}
                 & \lgc\drts{Number of neutrals}{Lower divertor, outer leg}& \lgc\drts{Number of neutrals}{Lower divertor, inner leg}
                 & \lgc\drts{Number of neutrals}{Core}
                 & \dgc\drts{Number of plasma particles}{Upper divertor, outer leg}& \dgc\drts{Number of plasma particles}{Upper divertor, inner leg}
                 & \dgc\drts{Number of plasma particles}{Lower divertor, outer leg}& \dgc\drts{Number of plasma particles}{Lower divertor, inner leg} 
                 & \dgc\drts{Number of plasma particles}{Core}
                 & \lgc\drts{Number of neutrals}{Upper pump}& \lgc\drts{Number of neutrals}{Lower pump}\\ \hline \hline
        \lgc\drt{Number of neutrals}{Upper divertor, outer leg}         &\lgc\grebul&\grebul&   ~   &\redbul&   ~   &\grebul&   ~   &   ~   &   ~   &   ~   &\grebul&   ~    \\ \hline
        \lgc\drt{Number of neutrals}{Upper divertor, inner leg}         &\grebul&\lgc\grebul&   ~   &   ~   &\redbul&\redbul&\grebul&\redbul&   ~   &\redbul&   ~   &   ~    \\ \hline
        \lgc\drt{Number of neutrals}{Lower divertor, outer leg}         &   ~   &\redbul&\lgc\grebul&\grebul&   ~   &   ~   &   ~   &\grebul&   ~   &   ~   &   ~   &\grebul \\ \hline
        \lgc\drt{Number of neutrals}{Lower divertor, inner leg}         &\redbul&\redbul&\grebul&\lgc\grebul&   ~   &   ~   &   ~   &\redbul&\grebul&   ~   &   ~   &   ~    \\ \hline
        \lgc\drt{Number of neutrals}{Core}                              &   ~   &   ~   &\redbul&\redbul&\lgc\grebul&\redbul&   ~   &   ~   &   ~   &\grebul&   ~   &   ~    \\ \hline
        \dgc\drt{Number of plasma particles}{Upper divertor, outer leg} &\grebul&\redbul&   ~   &   ~   &   ~   &\dgc\grebul&   ~   &   ~   &   ~   &   ~   &   ~   &   ~    \\ \hline
        \dgc\drt{Number of plasma particles}{Upper divertor, inner leg} &\redbul&\grebul&   ~   &   ~   &\redbul&\redbul&\dgc\grebul&   ~   &   ~   &   ~   &   ~   &   ~    \\ \hline
        \dgc\drt{Number of plasma particles}{Lower divertor, outer leg} &   ~   &   ~   &\grebul&\redbul&   ~   &   ~   &   ~   &\dgc\grebul&   ~   &   ~   &   ~   &   ~    \\ \hline
        \dgc\drt{Number of plasma particles}{Lower divertor, inner leg} &   ~   &\redbul&\redbul&\grebul&   ~   &   ~   &   ~   &\redbul&\dgc\grebul&   ~   &   ~   &   ~    \\ \hline
        \dgc\drt{Number of plasma particles}{Core}                      &   ~   &   ~   &   ~   &\redbul&\redbul&\redbar&\redbar&\redbar&\redbar&\dgc\grebul&   ~   &   ~    \\ \hline
        \lgc\drt{Number of neutrals}{Upper pump}                        &   ~   &   ~   &   ~   &   ~   &   ~   &   ~   &   ~   &   ~   &   ~   &   ~   &\lgc   ~   &   ~    \\ \hline
        \lgc\drt{Number of neutrals}{Lower pump}                        &\redbul&   ~   &   ~   &   ~   &   ~   &   ~   &   ~   &   ~   &   ~   &   ~   &   ~   &\lgc   ~    \\ \hline
    \end{tabular}}
\end{table}

\section{Future work} \label{sec:future}
In this article, we extended a specific network topology identification algorithm to include perturbation signals and applied it to a simulated tokamak model to identify its DSM. To get better insight into the results, we can perform Monte Carlo simulations, where in each run new input and disturbance signals are generated and the DSM is calculated. This leads to multiple results for the same network, which gives a more stable picture than just the result of a single simulation. Second, the input signals can be adapted to be more realistic. In the simulations in this article, the input signals are weighted prbs signals. In practice, the gas pellet injection to the core has indeed this kind of discrete behaviour, but the gas inlet flows to the divertor legs have continuous behaviour, for which multi-sines are commonly used for identification purposes. Performing simulations with these kind of signals leads to more insight as well. Especially, because the current input signals have a random characteristic, while multi-sine signals do not. Therefore, it is expected that the effect of the multi-sine inputs on the states can more easily be distinguished from the effect of the disturbances, leading to more correctly identified interconnections and thus to better results. 

To improve the algorithm, the method in \cite{Venkitaraman2020} can be explored and extended to included perturbations signals. This method is more simple as no hyperparameters are involved in the algorithm and therefore, it is less computational heavy, especially for large networks. In addition, the existence of an interconnection can be quantified by separating the data in several blocks (similar as in k-fold cross validation) and determining the topology for each subset of data. This leads to score between 0 and 1 for each edge and together with a threshold, the best network can be selected \cite{Baek2025}. The expectation is that this quantification is an indication of the importance of the interconnection, and it could also reflect the gain of the interconnections. 

Based on the foregoing discussion, we can also aim for identifying a static DSM from data instead of a binary DSM. A static DSM does not only represent the structure between the states, but also contains the gains of the interconnections. These gains can be represented by the relative gain array, which is used in exhaust control as well \cite{Figueiredo2025}. Even the full dynamics of the interconnections can be identified, leading to a dynamic DSM. Several algorithms are available for this of which a number have been implemented \cite{Sysdynet2026}. 

The potential of using topology identification for obtaining a DSM from data has been illustrated in the simulation example. One of the next steps is to apply this method to more advanced and realistic simulation models, such as JINTRAC, to investigate how this method extends to practice. In this simulation environment, we can directly obtain data of the signals of interest. However, in practice measurements are obtained via a set of diagnostics posing additional requirements on the available measurements. The mapping between the signals of interest and the diagnostics, referred to as the diagnostics model, can be obtained via a similar approach. This can be used to synthesize the optimal diagnostic set. In future work, we can extend the framework by including limit and state transitions. Finally, the DSM can serve as a basis to synthesize a supervisory controller for the PCS. 

\section{Conclusions} \label{sec:conclusion}
A DSM is a powerful tool to visualize, analyse, and manage complex systems. Constructing a DSM by interviewing experts can be time-consuming and heavily depends on the particular experts and their available knowledge. This entails the risk of incomplete information, which may result in an imbalanced DSM with missing and superfluous links. Based on the equivalence between a DSM and an adjacency matrix, we introduced the new idea to use dynamic network identification methods to obtain a DSM from data. This does not require understanding of the structure and behaviour of the system, only data of the signals of interest. To perform this identification, we extended a Bayesian selection model for topology identification to account for perturbation signals. By applying this algorithm to a simulated tokamak model of STEP, we illustrated that a DSM can be identified from data in this way. We showed that the algorithm can deal with disturbances that process through the system and that there still is some room for improving the method. In the nearby future, more insight into this algorithm can be gained by performing additional simulations. Other algorithms can be explored for obtaining a binary, static, or even dynamic DSM from data. Applying the method to more advanced simulation data will show how the promising results extend to practice and it will form the basis for application to integrated simulations such as JINTRAC. The framework can be extended and serve as a basis for synthesizing a supervisory controller for the PCS.


\bmsubsection*{Acknowledgements}
DIFFER is part of the institutes organization of NWO. 
This work has been funded by the Spherical Tokamak for Energy Production (STEP), a UKAEA programme to design and build a prototype fusion energy plant and a path to commercial fusion. 
This publication is part of the project Balls to the Wall (project no. 19695) of the research programme NWO Talent Programme VIDI, financed in part by the Dutch Research Council (NWO). 
This work has been carried out within the framework of the EUROfusion Consortium, funded by the European Union via the Euratom Research and Training Programme (Grant Agreement No 101052200 - EUROfusion).


\bmsubsection*{Conflicts of Interest}
The authors declare no conflicts of interest.

\bmsubsection*{Data availability}
Data will be made available on reasonable request. 

\bibliography{bibliography-wiley}


\end{document}